\documentclass[
aps,%
12pt,%
final,%
notitlepage,%
oneside,%
onecolumn,%
nobibnotes,%
nofootinbib,%
superscriptaddress,%
noshowpacs,%
centertags]%
{revtex4}

\usepackage{bm}
\usepackage{graphics}
\usepackage{rotating}
\usepackage{amsmath}
\usepackage{amssymb}
\usepackage{epsfig}

\begin{document}

\title{MASS SPECTRUM OF HEAVY TETRAQUARKS IN VARIATIONAL APPROACH}
\author{\firstname{A.~V.} \surname{Eskin}\footnote
{Talk presented at the session-conference "Physics of Fundamental Interactions" 
dedicated to the 70th anniversary of the birth of RAS Academician Valery Anatolyevich Rubakov 
17-21 February 2025, https://indico.inr.ac.ru/event/5/}
\affiliation{Samara University, Samara, Russia}}
\author{\firstname{A.~P.} \surname{Martynenko}}
\affiliation{Samara University, Samara, Russia}
\author{\firstname{F.~A.} \surname{Martynenko}}
\affiliation{Samara University, Samara, Russia}

\begin{abstract}
Within the framework of the quark model and the variational method, the bound states of four heavy quarks 
(tetraquarks) are investigated. The basis variational wave functions are chosen in the Gaussian form. 
The matrix elements of the Hamiltonian are calculated analytically. Numerical calculation of the energy 
levels of tetraquarks is performed using a program in the Matlab system. Hyperfine structure 
of the spectrum is calculated. To increase the accuracy of the calculation, relativistic corrections 
are taken into account.
\end{abstract}

\maketitle

\section{Introduction}

Studies of bound states of heavy quarks (mesons and baryons) have been carried out within 
the theory of strong interaction - quantum chromodynamics for many decades. During this time, 
various methods for the investigation of quark bound states based on quantum field principles have been formulated. 
At present, the greatest activity in studying both the hadron mass spectrum and various reactions of their 
formation and decay is observed within the framework of non-relativistic quantum chromodynamics, 
QCD sum rules and various versions of quark models.

Along with the usual quark-antiquark and three-quark hadrons, exotic hadrons built of four, five quarks 
and antiquarks (tetraquarks, pentaquarks, etc.) have also been studied in constructing the quark model. 
The investigation of characteristics of such multiquark states improves our understanding of the properties 
of strong interaction, including non-perturbative aspects of QCD. The first candidate for tetraquarks 
was discovered by the Belle collaboration in 2003 \cite{belle}, and since then, dozens of more candidates 
for exotic hadron states have been discovered \cite{brodsky,lebed}. It can be said that a qualitatively 
new period has begun in the study of bound states of quarks and gluons. In 2020, the LHCb collaboration 
discovered a new resonant state with a mass of about 6.9 GeV \cite{lhcb}. This state was later confirmed 
by other collaborations, such as ATLAS \cite{atlas} and CMS \cite{cms,cms1}, and was considered a candidate 
for a tetraquark consisting of two $c$ quarks and two $\bar c$ antiquarks. The properties of this state, 
including the mechanisms of its production and decay, are currently being studied.

Since the quantum problem with the number of particles greater than or equal to three does not have 
an exact solution, various models \cite{efg1,efg2,efg3,t11,t12,t13,m1,m2,tc1,tc2,tc3,t2,t4} are used 
to calculate the observed characteristics 
of multiparticle states in QCD (see other references in \cite{lebed,efg3}). For example, the model 
of quark-diquark interaction in a baryon, or the model 
of diquark-antidiquark interaction in a tetraquark, have become widely known. The numerical results 
of calculations obtained within its framework are in good agreement with already known experimental data, 
and also have predictive power for states that have not yet been discovered. Many results obtained 
in the relativistic quark model \cite{efg1,efg2,efg3} subsequently had good experimental confirmation.

This paper is devoted to specific hadron states - tetraquarks $(cc\bar c\bar c)$, $(bb\bar b\bar b)$,
$(cc\bar b\bar b)$ consisting of heavy quarks $c$ and $b$. 
To calculate the mass spectrum of tetraquarks, we develop a variational method for the case of four particles, 
which was previously used to study various three-particle Coulomb systems in QED \cite{p1,p2,p3,v1,v2,v3}. 
The variational method is one of the most commonly used approaches to solving quantum mechanical problems with 
many particles. Although it only provides an approximate solution to the problem when calculating the energy 
levels of the bound state, the accuracy of such calculations is extremely high.
Thus, we do not use the diquark-antidiquark model and consider the interaction of all four particles 
inside the tetraquark.

\section{Variational method in the tetraquark problem}

Let us consider the description of bound states of four particles within the variational 
method in the coordinate representation. Let ${\bf r}_1$, ${\bf r}_2$, ${\bf r}_3$, ${\bf r}_4$ be the 
radius vectors of the particles in the initial reference frame. It is convenient to move on to the Jacobi 
coordinates $\boldsymbol\rho$, $\boldsymbol\lambda$, $\boldsymbol\sigma$, which are related to the initial 
coordinates as follows:
\begin{eqnarray}
{\bf r}_1=-\frac{m_2}{m_{12}}\boldsymbol\rho+\frac{m_{34}}{m_{1234}}\boldsymbol\sigma,~~~
{\bf r}_2=\frac{m_1}{m_{12}}\boldsymbol\rho+\frac{m_{34}}{m_{1234}}\boldsymbol\sigma,\\
{\bf r}_3=-\frac{m_4}{m_{34}}\boldsymbol\lambda-\frac{m_{12}}{m_{1234}}\boldsymbol\sigma,~~~
{\bf r}_4=\frac{m_3}{m_{34}}\boldsymbol\lambda-\frac{m_{12}}{m_{1234}}\boldsymbol\sigma,
\end{eqnarray}
where $m_i$ is the mass of ith particle, $m_{ij}=m_i+m_j$, $m_{1234}=\sum_{i=1}^4 m_i$.

\begin{figure}[htbp]
\centering
\includegraphics[scale=0.7]{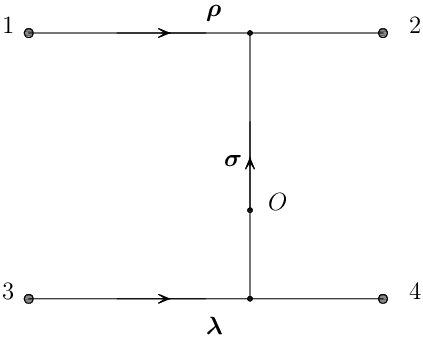}
\caption{The Jacobi coordinates in a four-particle system.}
\label{ris1}
\end{figure}

This choice of coordinates is shown in Fig.~\ref{ris1}. The coordinate $\boldsymbol\rho$ represents 
the relative distance between the first pair of particles 1 and 2. The coordinate $\boldsymbol\lambda$ 
represents the relative distance between the second pair of particles 3 and 4. The coordinate 
$\boldsymbol\sigma$ represents the relative distance between the centers of mass of first and 
second pair.

In the Jacobi coordinates, the kinetic energy operator of the system has the form:
\begin{equation}
\label{f1}
\hat T=\frac{{\bf p}^2_\rho}{2\mu_1}+\frac{{\bf p}^2_\lambda}{2\mu_2}+\frac{{\bf p}^2_\sigma}{2\mu_3},
\end{equation}
where the reduced masses $\mu_1$, $\mu_2$, $\mu_3$ are equal to:
\begin{equation}
\label{f2}
\mu_1=\frac{m_1 m_2}{m_{12}},~~~\mu_2=\frac{m_3 m_4}{m_{34}},~~~\mu_3=\frac{m_{12} m_{34}}{m_{1234}}.
\end{equation}

To find the wave function of the system, either the exponential or the Gaussian form is usually used
\cite{v1,v2,v3}. We choose variational functions of the ground state of four-particle system in the 
Gaussian form as in previous works for three-particle systems \cite{p1,p2,p3}:
\begin{equation}
\label{f3}
\Psi(\boldsymbol\rho,\boldsymbol\lambda,\boldsymbol\sigma)=\sum_{I=1}^K
C_I e^{-\frac{1}{2}\bigl[A_{11}(I)
\boldsymbol\rho^2+
2A_{12}(I)\boldsymbol\rho\boldsymbol\lambda+A_{22}(I)\boldsymbol\lambda^2+
2A_{13}(I)\boldsymbol\rho\boldsymbol\sigma+2A_{23}(I)\boldsymbol\lambda\boldsymbol\sigma+
A_{33}(I)\boldsymbol\sigma^2\bigr]},
\end{equation}
where $A_{ij}(I)$ is the matrix of nonlinear variational parameters, $C_I$ are linear variational 
parameters.

The normalization condition includes the following integral:
\begin{equation}
\label{f4}
B=\int d{\boldsymbol\rho}\int d{\boldsymbol\lambda}\int d{\boldsymbol\sigma}
e^{-\frac{1}{2}(B_{11}
\boldsymbol\rho^2+
2B_{12}\boldsymbol\rho\boldsymbol\lambda+B_{22}\boldsymbol\lambda^2+
2B_{13}\boldsymbol\rho\boldsymbol\sigma+2B_{23}\boldsymbol\lambda\boldsymbol\sigma+
B_{33}\boldsymbol\sigma^2)}=
\frac{(8\pi^3)^{3/2}}{B_{33}^{3/2}(\tilde B_{11}\tilde B_{22}-\tilde B_{12}^2)^{3/2}},
\end{equation}
where the notations for the parameters $\tilde B_{ij}$ are introduced:
\begin{equation}
\label{f5}
\tilde B_{11}=B_{11}-\frac{B_{13}^2}{B_{33}},~~~\tilde B_{22}=B_{22}-\frac{B_{23}^2}{B_{33}},~~~
\tilde B_{12}=B_{12}-\frac{B_{13}B_{23}}{B_{33}},~~~B_{kl}=A_{kl}(I)+A_{kl}(J).
\end{equation}

As a result, the normalization coefficient in wave function of a system in the coordinate 
representation has the form:
\begin{equation}
\label{f5a}
{\cal N}=\sum_{I=1}^K\sum_{J=1}^K C_I C_J
\frac{(8\pi^3)^{3/2}}{B_{33}^{3/2}(I,J)(\tilde B_{11}(I,J)\tilde B_{22}(I,J)-
\tilde B_{12}^2(I,J))^{3/2}},
\end{equation}
where in formula \eqref{f5a} the dependence of all matrix elements on two indices 
is emphasized: $B_{kl}(I,J)=A_{kl}(I)+A_{kl}(J)$.

In what follows we consider tetraquarks consisting of $c$ and $b$ quarks. For the tetraquark 
$(cc\bar c\bar c)$, $(bb\bar b\bar b)$, the symmetrization of coordinate wave function is performed, 
which is connected with the permutation of identical quarks or antiquarks. The coordinate part 
of total wave function must be symmetric under the replacement $\boldsymbol\rho\to -\boldsymbol\rho$, $\boldsymbol\lambda\to -\boldsymbol\lambda$. In order to obtain a wave function \eqref{f20} that 
is symmetrical under such a change of coordinates, it is necessary to add to the exponent in \eqref{f20} 
three more exponents that differ in the signs of the nonlinear variational parameters: 
$A_{12}(I)\to -A_{12}(I)$, $A_{13}(I)\to -A_{13}(I)$;
$A_{12}(I)\to -A_{12}(I)$, $A_{23}(I)\to -A_{23}(I)$; $A_{13}(I)\to -A_{13}(I)$, $A_{23}(I)\to -A_{23}(I)$. 
The matrix elements themselves are determined by general formulas that we present below, 
with the corresponding replacements in variational parameters.

The spin wave function of the tetraquark $\chi_{SS_z}^{S_{12}S_{34}}$ is also symmetric under the permutation 
of quarks 1 and 2 and antiquarks 3 and 4 ($S_{12}=1$ is the spin of a pair of quarks, $S_{34}=1$ 
is the spin of a pair of antiquarks):
\begin{eqnarray}
\label{f5b}
\chi_{00}^{11}=\frac{1}{\sqrt{12}}\left(2\uparrow\uparrow\downarrow\downarrow-
\uparrow \downarrow   \uparrow \downarrow - \uparrow  \downarrow\downarrow \uparrow-
\downarrow\uparrow\uparrow\downarrow-\downarrow \uparrow \downarrow\uparrow+
2 \downarrow\downarrow \uparrow\uparrow\right),  \\
\chi_{11}^{11}=\frac{1}{2}\left(\uparrow\uparrow\uparrow\downarrow+ \uparrow\uparrow  \downarrow\uparrow-
\uparrow \downarrow \uparrow\uparrow-\downarrow\uparrow\uparrow\uparrow\right),  \\
\chi_{22}^{11}=\uparrow\uparrow\uparrow\uparrow,~~~\uparrow={1\choose{0}},~~~\downarrow={0\choose{1}}.
\end{eqnarray}

The permutation of quarks 1 and 2, as well as antiquarks 3 and 4, must change the sign of total 
wave function. Therefore, color part of wave function of a pair of quarks and a pair 
of antiquarks is antisymmetric.

We also calculate the matrix elements of kinetic and potential energy, which are necessary for solving 
the matrix eigenvalue problem. The kinetic energy operator consists of three terms, which are determined 
by the Laplace operators for the variables $\boldsymbol\rho$, $\boldsymbol\lambda$,
$\boldsymbol\sigma$. All these matrix elements are calculated analytically in the basis 
of the Gaussian functions. The calculation results are as follows:
\begin{equation}
\label{f6}
<\Delta_\rho>=\frac{48\sqrt{2}\pi^{9/2}}{det B^{5/2}}
\Bigl[A_{11}^2(B_{22}B_{33}-B_{23}^2)+A_{12}^2(B_{11}B_{33}-B_{13}^2)+
A_{13}^2(B_{11}B_{22}-B_{12}^2)-A_{11}detB+
\end{equation}
\begin{displaymath}
2A_{11}A_{12}(B_{13}B_{23}-B_{12}B_{33})+2A_{11}A_{13}(B_{12}B_{23}-B_{13}B_{22})+
2A_{12}A_{13}(B_{12}B_{13}-B_{11}B_{23})\Bigr],
\end{displaymath}
\begin{equation}
\label{f7}
<\Delta_\lambda>=\frac{48\sqrt{2}\pi^{9/2}}{det B^{5/2}}
\Bigl[A_{12}^2(B_{22}B_{33}-B_{23}^2)+A_{22}^2(B_{11}B_{33}-B_{13}^2)+
A_{23}^2(B_{11}B_{22}-B_{12}^2)-A_{22}detB+
\end{equation}
\begin{displaymath}
2A_{22}A_{12}(B_{13}B_{23}-B_{12}B_{33})+2A_{22}A_{23}(B_{12}B_{13}-B_{11}B_{23})+
2A_{12}A_{23}(B_{12}B_{23}-B_{13}B_{22})\Bigr],
\end{displaymath}
\begin{equation}
\label{f8}
<\Delta_\sigma>=\frac{48\sqrt{2}\pi^{9/2}}{det B^{5/2}}
\Bigl[A_{13}^2(B_{22}B_{33}-B_{23}^2)+A_{23}^2(B_{11}B_{33}-B_{13}^2)+
A_{33}^2(B_{11}B_{22}-B_{12}^2)-A_{33}detB+
\end{equation}
\begin{displaymath}
2A_{13}A_{23}(B_{13}B_{23}-B_{12}B_{33})+2A_{33}A_{13}(B_{12}B_{23}-B_{13}B_{22})+
2A_{33}A_{23}(B_{12}B_{13}-B_{11}B_{23})\Bigr],
\end{displaymath}
where the result is expressed in terms of original nonlinear parameters,
\begin{equation}
\label{f9}
det B=B_{11}B_{22}B_{33}-B_{11} B_{23}^2-B_{12}^2B_{33}+2 B_{12}B_{13} B_{23}- B_{13}^2 B_{22}.
\end{equation}

The matrix elements of potential energy consist in non-relativistic approximation 
of the matrix elements of the paired Coulomb terms and the matrix elements of the paired confinement 
potentials. Direct calculation of integrals with wave functions \eqref{f3} gives the following results:
\begin{equation}
\label{f12}
<\frac{1}{|{\bf r}_2-{\bf r}_1|}>=<\frac{1}{\rho}>=\frac{32\pi^4}{B_{33}^{3/2}
\sqrt{\tilde B_{22}}(\tilde B_{11}\tilde B_{22}-\tilde B_{12}^2)},
\end{equation}
\begin{equation}
\label{f13}
<\frac{1}{|{\bf r}_4-{\bf r}_3|}>=<\frac{1}{\lambda}>=\frac{32\pi^4}{B_{33}^{3/2}
\sqrt{\tilde B_{11}}(\tilde B_{11}\tilde B_{22}-\tilde B_{12}^2)},
\end{equation}
\begin{equation}
\label{f14}
<\frac{1}{|{\bf r}_1-{\bf r}_3|}>=
<\frac{1}{|\boldsymbol\sigma+\frac{m_4}{m_{34}}\boldsymbol\lambda-
\frac{m_2}{m_{12}}\boldsymbol\rho|}>=\frac{32\pi^4}{F_{1}^{3/2}
\sqrt{F_2-\frac{F_3^2}{F_1}}
[(F_2-\frac{F_3^2}{F_1})(B_{33}-\frac{F_4^2}{F_1})-(F_5-\frac{F_3F_4}{F_1})^2]}.
\end{equation}
The result of calculating the matrix element \eqref{f14} is expressed in terms of quantities 
that arise in the process of replacing the integration variables:
\begin{equation}
\label{f15}
F_1=B_{11}+B_{33}\frac{m_2^2}{m_{12}^2}+2B_{13}\frac{m_2}{m_{12}},
\end{equation}
\begin{equation}
\label{f16}
F_2=B_{22}+B_{33}\frac{m_4^2}{m_{34}^2}-2B_{23}\frac{m_4}{m_{34}},
\end{equation}
\begin{equation}
\label{f17}
F_3=B_{12}+B_{23}\frac{m_2}{m_{12}}-B_{33}\frac{m_2m_4}{m_{12}m_{34}}-B_{13}\frac{m_4}{m_{34}},
\end{equation}
\begin{equation}
\label{f18}
F_4=B_{13}+B_{33}\frac{m_2}{m_{12}},~~~F_5=B_{23}-B_{33}\frac{m_4}{m_{34}}.
\end{equation}

Three remaining matrix elements with Coulomb interaction can be obtained from \eqref{f14} 
by simple parameter changes:
\begin{equation}
\label{f19}
<\frac{1}{|{\bf r}_1-{\bf r}_4|}>=<\frac{1}{|\boldsymbol\sigma-\frac{m_3}{m_{34}}\boldsymbol\lambda-
\frac{m_2}{m_{12}}\boldsymbol\rho|}>,~~~\frac{m_4}{m_{34}}\to-\frac{m_3}{m_{34}},
\end{equation}
\begin{equation}
\label{f20}
<\frac{1}{|{\bf r}_2-{\bf r}_3|}>=<\frac{1}{|\boldsymbol\sigma+\frac{m_1}{m_{12}}\boldsymbol\rho-
\frac{m_4}{m_{34}}\boldsymbol\lambda|}>,~~~\frac{m_2}{m_{12}}\to-\frac{m_1}{m_{12}},
\end{equation}
\begin{equation}
\label{f21}
<\frac{1}{|{\bf r}_2-{\bf r}_4|}>=<\frac{1}{|\boldsymbol\sigma+\frac{m_1}{m_{12}}\boldsymbol\rho-
\frac{m_3}{m_{34}}\boldsymbol\lambda|}>,~~~\frac{m_2}{m_{12}}\to-\frac{m_1}{m_{12}},~~~
\frac{m_4}{m_{34}}\to-\frac{m_3}{m_{34}}.
\end{equation}

In the case of confinement potential, we assume that it consists of pair potentials 
of quark-quark or quark-antiquark interactions. Therefore, the corresponding matrix elements 
have the form:
\begin{equation}
\label{f22}
<|{\bf r}_2-{\bf r}_1|>=<\rho>=\frac{64\pi^4\sqrt{\tilde B_{22}}}
{B_{33}^{3/2}(\tilde B_{11}\tilde B_{22}-\tilde B_{12}^2)^2},
\end{equation}
\begin{equation}
\label{f23}
<|{\bf r}_3-{\bf r}_4|>=<\lambda>=\frac{64\pi^4\sqrt{\tilde B_{11}}}
{B_{33}^{3/2}(\tilde B_{11}\tilde B_{22}-\tilde B_{12}^2)^2},
\end{equation}
\begin{equation}
\label{f24}
|{\bf r}_1-{\bf r}_3|>=
<|\boldsymbol\sigma+\frac{m_4}{m_{34}}\boldsymbol\lambda-\frac{m_2}{m_{12}}\boldsymbol\rho|>=
\frac{64\pi^4\sqrt{F_2-\frac{F_3^2}{F_1}}}{F_{1}^{3/2}
[(F_2-\frac{F_3^2}{F_1})(B_{33}-\frac{F_4^2}{F_1})-(F_5-\frac{F_3F_4}{F_1})^2]^2}.
\end{equation}

Remaining confinement matrix elements are calculated using \eqref{f24} and parameter changes 
that are analogous to \eqref{f19}-\eqref{f21}:
\begin{equation}
\label{f25}
<|{\bf r}_1-{\bf r}_4|>=
<|\boldsymbol\sigma-\frac{m_3}{m_{34}}\boldsymbol\lambda-\frac{m_2}{m_{12}}\boldsymbol\rho|>,~~~
\frac{m_4}{m_{34}}\to-\frac{m_3}{m_{34}},
\end{equation}
\begin{equation}
\label{f26}
<|{\bf r}_2-{\bf r}_3|>=
<|\boldsymbol\sigma+\frac{m_4}{m_{34}}\boldsymbol\lambda+\frac{m_1}{m_{12}}\boldsymbol\rho|>,~~~
\frac{m_2}{m_{12}}\to-\frac{m_1}{m_{12}},
\end{equation}
\begin{equation}
\label{f27}
<|{\bf r}_2-{\bf r}_4|>=
<|\boldsymbol\sigma-\frac{m_3}{m_{34}}\boldsymbol\lambda+\frac{m_1}{m_{12}}\boldsymbol\rho|>,~~~
\frac{m_2}{m_{12}}\to-\frac{m_1}{m_{12}},~~~\frac{m_4}{m_{34}}\to-\frac{m_3}{m_{34}}.
\end{equation}
 
Since this calculation is performed for heavy tetraquarks consisting of $b$ and $c$ quarks and antiquarks, 
it is convenient to specify the quark masses in the input file directly in GeV. 
Total Coulomb potential for four particles is determined by the following sum of terms:
\begin{equation}
\label{f27a}
\Delta V^C(\boldsymbol\rho,\boldsymbol\lambda,\boldsymbol\sigma)=
\Bigl[
-\frac{2\alpha_{s12}}{3\rho}-\frac{2\alpha_{s34}}{3\lambda}-
\frac{4\alpha_{s13}}{3}\frac{1}{|\boldsymbol\sigma+\frac{m_4}{m_{34}}
\boldsymbol\lambda-\frac{m_2}{m_{12}}\boldsymbol\rho|}-
\frac{4\alpha_{s14}}{3}\frac{1}{|\boldsymbol\sigma-\frac{m_3}{m_{34}}
\boldsymbol\lambda-\frac{m_2}{m_{12}}\boldsymbol\rho|}-
\end{equation}
\begin{displaymath}
\frac{4\alpha_{s23}}{3}\frac{1}{|\boldsymbol\sigma+\frac{m_4}{m_{34}}
\boldsymbol\lambda+\frac{m_1}{m_{12}}\boldsymbol\rho|}-
\frac{4\alpha_{s24}}{3}\frac{1}{|\boldsymbol\sigma-\frac{m_3}{m_{34}}
\boldsymbol\lambda+\frac{m_1}{m_{12}}\boldsymbol\rho|}
\Bigr],
\end{displaymath}
where $\alpha_{s~ij}$ is the constant of strong interaction for a pair of particles $i$ and $j$. 
This type of potential means that particles 1 and 2, 3 and 4 are heavy quarks or
antiquarks (the quark-quark potential contains an additional factor of $1/2$ compared 
to the potential of quark-antiquark interaction).

Total confinement potential for four particles is determined by the sum of two-particle 
potentials in the form:
\begin{eqnarray}
\label{f27b}
\Delta V^{conf}(\boldsymbol\rho,\boldsymbol\lambda,\boldsymbol\sigma)=
\Bigl[
\frac{1}{2} A|{\boldsymbol\rho}|+\frac{1}{2} A|{\boldsymbol\lambda}|+
 A|\boldsymbol\sigma+\frac{m_4}{m_{34}}\boldsymbol\lambda-\frac{m_2}{m_{12}}
\boldsymbol\rho|+\\
 A|\boldsymbol\sigma-\frac{m_3}{m_{34}}\boldsymbol\lambda-\frac{m_2}{m_{12}}
\boldsymbol\rho|+
 A|\boldsymbol\sigma+\frac{m_4}{m_{34}}\boldsymbol\lambda+\frac{m_1}{m_{12}}
\boldsymbol\rho|+
 A|\boldsymbol\sigma-\frac{m_3}{m_{34}}
\boldsymbol\lambda+\frac{m_1}{m_{12}}\boldsymbol\rho|+ B
\Bigr], \nonumber
\end{eqnarray}
where the parameters in confinement potential \eqref{f27b} and in the Coulomb potential were chosen 
in numerical calculations in exactly the same form as for the pair interaction of quarks and antiquarks 
in mesons and baryons:
$A=0.18~GeV^2$, $\alpha_{s~cc}=0.314$ (for $c$-quarks), $\alpha_{s~bb}=0.207$ (for $b$-quarks),
$\alpha_{s~cb}=0.264$ (for $c,b$-quarks), $m_c=1.55~GeV$, $m_b=4.88~GeV$.
Total value of the energy spectrum shift constant was chosen to be $B=-0.8~GeV$.

Along with the coordinate representation, the momentum representation is also often used. 
The wave function of a four-particle system can be obtained by calculating the Fourier transform 
\eqref{f3} in the form:
\begin{equation}
\label{f3a}
\Psi({\bf p},{\bf q},{\bf k})=\sum_{I=1}^K\frac{C_I(2\pi)^{\frac{9}{2}}}
{\sqrt{{\cal N}}(det A)^{3/2}}
e^{-\frac{1}{2det A}[{\bf p}^2(A_{22}A_{33}-
A_{23}^2)+{\bf q}^2(A_{11}A_{33}-A_{13}^2)+{\bf k}^2(A_{11}A_{22}-A_{12}^2)]}
\end{equation}
\begin{displaymath}
e^{-\frac{1}{2det A}[2{\bf p}{\bf k}(A_{12}A_{23}-A_{13}A_{22})+
2{\bf q}{\bf k}(A_{12}A_{13}-A_{11}A_{23})+2{\bf p}{\bf q}(A_{13}A_{23}-A_{12}A_{33})]},
\end{displaymath}
where ${\bf p}$, ${\bf q}$, ${\bf k}$ are relative momenta for particles 12, 34 and the pair 
of particles 12 and 34, and all matrix elements $A_{kl}=A_{kl}(I)$.

The normalization factor of wave function \eqref{f3a}, which is obtained after calculating 
the integrals over relative momenta, is equal to 1:
\begin{equation}
\label{f3b}
<\Psi|\Psi>=
\int|\Psi({\bf p},{\bf q},{\bf k})|^2\frac{d{\bf p} d{\bf q} d{\bf k}}{(2\pi)^9}=
\sum_{I=1}^K\sum_{J=1}^K \frac{C_I C_J 16\sqrt{2}\pi^{\frac{9}{2}}}
{{\cal N}}
\frac{1}{det Q(I,J)^{\frac{3}{2}}}\frac{1}{det A(I)^{\frac{3}{2}}}
\frac{1}{det A(J)^{\frac{3}{2}}}.
\end{equation}

The elements of the matrix $Q$ in \eqref{f3b} are expressed in terms of nonlinear parameters:

\begin{equation}
\label{f3c}
Q_{11}=\frac{(A_{22}(I)A_{33}(I)-A^2_{23}(I))}{det A(I)}+
\frac{(A_{22}(J)A_{33}(J)-A_{23}(J)^2)}{det A(J)},
\end{equation}
\begin{equation}
Q_{22}=\frac{(A_{11}(I)A_{33}(I)-A^2_{13}(I))}{det A(I)}+
\frac{(A_{11}(J)A_{33}(J)-A^2_{13}(J))}{det A(J)},
\end{equation}
\begin{equation}
Q_{33}=\frac{(A_{11}(I)A_{22}(I)-A^2_{12}(I))}{det A(I)}+
\frac{(A_{11}(J)A_{22}(J)-A^2_{12}(J))}{det A(J)},
\end{equation}
\begin{equation}
Q_{13}=\frac{(A_{12}(I)A_{23}(I)-A_{13}(I)A_{22}(I))}{det A(I)}+
\frac{(A_{12}(J)A_{23}(J)-A_{13}(J)A_{22}(J))}{det A(J)},
\end{equation}
\begin{equation}
Q_{23}=\frac{(A_{12}(I)A_{13}(I)-A_{11}(I)A_{23}(I))}{det A(I)}+
\frac{(A_{12}(J)A_{13}(J)-A_{11}(J)A_{23}(J))}{det A(J)},
\end{equation}
\begin{equation}
Q_{12}=\frac{(A_{13}(I)A_{23}(I)-A_{12}(I)A_{33}(I))}{det A(I)}+
\frac{(A_{13}(J)A_{23}(J)-A_{12}(J)A_{33}(J))}{det A(J)}.
\end{equation}

In connection with the discovery of numerous new hadron states, calculations of the cross sections 
of tetraquark production in various reactions have become especially relevant \cite{s7,s8,s9,s10,tt}. Since the amplitude 
of the hadron production process is represented as a convolution of the amplitude of free quark production 
with the wave functions of bound states, calculating the tetraquark wave function is an important task. 
In the nonrelativistic approximation, the cross section of tetraquark production is determined 
by the value of the wave function at zero relative coordinates:
\begin{equation}
\label{f3d}
\Psi(0,0,0)=\int\Psi({\bf p},{\bf q},{\bf k})
\frac{d{\bf p} d{\bf q} d{\bf k}}{(2\pi)^9}=
\sum_{I=1}^K \frac{C_I }{\sqrt{{\cal N}}}
\frac{1}{det \tilde Q(I)^{\frac{3}{2}}}\frac{1}{det A(I)^{\frac{3}{2}}},
\end{equation}
where the matrix elements $\tilde Q$ are determined by the first half of the terms 
in the matrix elements $Q$.

An increase in the accuracy of the calculation of the production cross sections can be achieved by taking 
into account various corrections to the non-relativistic approximation, including relativistic corrections 
determined by the relative momenta in the four-particle system.
Relativistic corrections in the production of bound states of four particles are determined by momentum 
integrals of the following form:
\begin{equation}
\label{f3e}
\langle\frac{{\bf p}^2}{m^2}\rangle=
\int\Psi({\bf p},{\bf q},{\bf k})\frac{{\bf p}^2}{m^2}
\frac{d{\bf p} d{\bf q} d{\bf k}}{(2\pi)^9}=
\sum_{I=1}^K \frac{3C_I}{\sqrt{{\cal N}}}
\frac{(\tilde Q_{22}\tilde Q_{33}-\tilde Q^2_{23})}
{det A(I)^{\frac{3}{2}}det \tilde Q(I)^{\frac{5}{2}}m^2}\times
\end{equation}
\begin{displaymath}
\Bigl[Erf\bigl(\sqrt{\frac{m^2 det\tilde Q}{2(\tilde Q_{22}\tilde Q_{33}-\tilde Q^2_{23}}}\bigr)-
\frac{\sqrt{2}}{3\sqrt{\pi}}\frac{m^3 (det\tilde Q)^{3/2}}{(\tilde Q_{22}\tilde Q_{33}-\tilde Q_{23}^2)^{3/2}}
\bigl(1+\frac{3(\tilde Q_{22}\tilde Q_{33}-\tilde Q_{23}^2)}{m^2 det\tilde Q}\bigr)
e^{-\frac{m^2 det\tilde Q}{2(\tilde Q_{22}\tilde Q_{33}-\tilde Q_{23}^2)}}
\Bigr],
\end{displaymath}
\begin{equation}
\label{f3f}
\langle\frac{{\bf q}^2}{m^2}\rangle=
\int\Psi({\bf p},{\bf q},{\bf k})\frac{{\bf q}^2}{m^2}
\frac{d{\bf p} d{\bf q} d{\bf k}}{(2\pi)^9}=
\sum_{I=1}^K \frac{3C_I}{\sqrt{{\cal N}}}
\frac{(\tilde Q_{11}\tilde Q_{33}-\tilde Q^2_{13})}
{det A(I)^{\frac{3}{2}}det \tilde Q(I)^{\frac{5}{2}}m^2}\times
\end{equation}
\begin{displaymath}
\Bigl[Erf\bigl(\sqrt{\frac{m^2 det\tilde Q}{2(\tilde Q_{11}\tilde Q_{33}-\tilde Q^2_{13}}}\bigr)-
\frac{\sqrt{2}}{3\sqrt{\pi}}\frac{m^3 (det\tilde Q)^{3/2}}{(\tilde Q_{11}\tilde Q_{33}-\tilde Q_{13}^2)^{3/2}}
\bigl(1+\frac{3(\tilde Q_{11}\tilde Q_{33}-\tilde Q_{13}^2)}{m^2 det\tilde Q}\bigr)
e^{-\frac{m^2 det\tilde Q}{2(\tilde Q_{11}\tilde Q_{33}-\tilde Q_{13}^2)}}
\Bigr],
\end{displaymath}
\begin{equation}
\label{f3g}
\langle\frac{{\bf k}^2}{m^2}\rangle=
\int\Psi({\bf p},{\bf q},{\bf k})\frac{{\bf k}^2}{m^2}
\frac{d{\bf p} d{\bf q} d{\bf k}}{(2\pi)^9}=
\sum_{I=1}^K \frac{3C_I}{\sqrt{{\cal N}}}
\frac{(\tilde Q_{11}\tilde Q_{22}-\tilde Q^2_{12})}
{det A(I)^{\frac{3}{2}}det \tilde Q(I)^{\frac{5}{2}}m^2}\times
\end{equation}
\begin{displaymath}
\Bigl[Erf\bigl(\sqrt{\frac{m^2 det\tilde Q}{2(\tilde Q_{11}\tilde Q_{22}-\tilde Q^2_{12}}}\bigr)-
\frac{\sqrt{2}}{3\sqrt{\pi}}\frac{m^3 (det\tilde Q)^{3/2}}{(\tilde Q_{11}\tilde Q_{22}-\tilde Q_{12}^2)^{3/2}}
\bigl(1+\frac{3(\tilde Q_{11}\tilde Q_{22}-\tilde Q_{12}^2)}{m^2 det\tilde Q}\bigr)
e^{-\frac{m^2 det\tilde Q}{2(\tilde Q_{11}\tilde Q_{22}-\tilde Q_{12}^2)}}
\Bigr],
\end{displaymath}
where Erf is the error function, $m$ is the mass of heavy quark in the case of a tetraquark $(QQ\bar Q\bar Q)$.
Despite the convergence of these integrals, when calculating them in the region of large values 
of relative momenta, one integral is cut off at the relativistic momentum $m$, since in the region 
of relativistic momenta we have poor knowledge of the wave function of the bound state.
Other relativistic corrections, which are determined by $\langle\frac{{\bf p}{\bf k}}{m^2}\rangle$,
$\langle\frac{{\bf q}{\bf k}}{m^2}\rangle$, $\langle\frac{{\bf p}{\bf q}}{m^2}\rangle$, 
are not given here, since their value turns out to be two orders of magnitude smaller 
than those considered \eqref{f3e}-\eqref{f3g}.

After calculating the wave functions of the tetraquarks, we can use them to analyze the motion of the 
quarks in the tetraquark. For this purpose, the radial distributions of relative momenta are calculated, 
which are determined by the following formulas:
\begin{equation}
\label{f4a}
W_1(p^2)=\sum_{I,J=1}^K \frac{C_I C_J}{{\cal N} det A(I)^{3/2}det A(J)^{3/2}}
\frac{8\pi^3 p^2}{(Q_{22}Q_{33}-Q_{23}^2)^{3/2}}e^{-\frac{det Q}{2(Q_{22}Q_{33}-Q_{23}^2)}p^2},
\end{equation}
\begin{equation}
\label{f4b}
W_2(q^2)=\sum_{I,J=1}^K \frac{C_I C_J}{{\cal N} det A(I)^{3/2}det A(J)^{3/2}}
\frac{8\pi^3 q^2}{(Q_{11}Q_{33}-Q_{13}^2)^{3/2}}e^{-\frac{det Q}{2(Q_{11}Q_{33}-Q_{13}^2)}q^2},
\end{equation}
\begin{equation}
\label{f4c}
W_3(k^2)=\sum_{I,J=1}^K \frac{C_I C_J}{{\cal N} det A(I)^{3/2}det A(J)^{3/2}}
\frac{8\pi^3 k^2}{(Q_{11}Q_{22}-Q_{12}^2)^{3/2}}e^{-\frac{det Q}{2(Q_{11}Q_{22}-Q_{12}^2)}k^2}.
\end{equation}

\begin{figure}[htbp]
\centering
\includegraphics[scale=0.55]{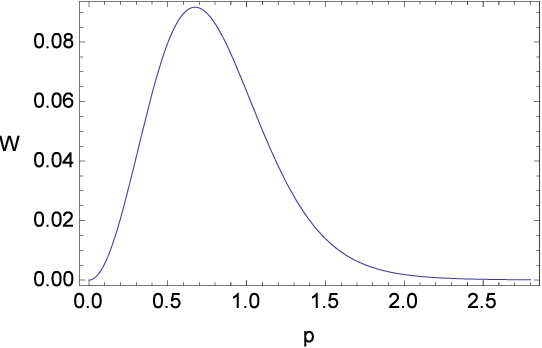}
\includegraphics[scale=0.55]{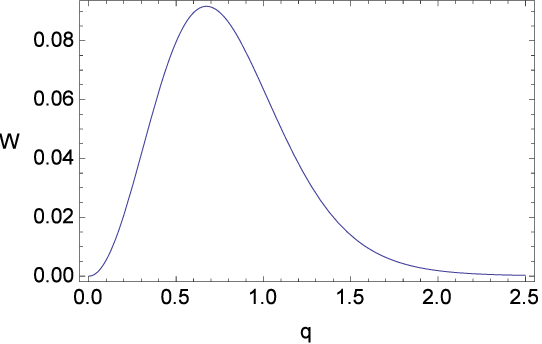}
\includegraphics[scale=0.55]{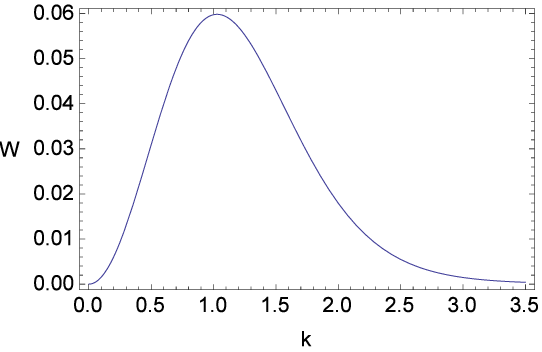}
\caption{Radial distributions of relative momenta $p$, $q$, $k$.}
\label{ris2}
\end{figure}

The plots of these distributions are shown in Fig.~\ref{ris2}. The functions $W_1(p^2)$, 
$W_2(q^2)$, $W_3(k^2)$ allow us to identify the characteristic momenta of relative motions 
of heavy quarks in a tetraquark.
The characteristic values of the relative momenta $p$ and $q$ in a tetraquark $(cc\bar c\bar c)$ are the same, 
and the characteristic value of the relative momentum $k$ is approximately twice as large.

When calculating the energy levels of the tetraquark, their hyperfine interaction was also taken into account.
Hyperfine structure of the tetraquark spectrum is determined by the pair potentials 
of spin-spin interaction in the ground state in the form:
\begin{equation}
\label{hfs1}
\Delta V^{hfs}({\boldsymbol\rho},{\boldsymbol\lambda},{\boldsymbol\sigma})=
a_{12}({\bf s}_1{\bf s}_2)+a_{34}({\bf s}_3{\bf s}_4)+a_{13}({\bf s}_1{\bf s}_3)+
a_{14}({\bf s}_1{\bf s}_4)+a_{23}({\bf s}_2{\bf s}_3)+a_{24}({\bf s}_2{\bf s}_4),
\end{equation}
\begin{eqnarray}
\label{hfs2}
a_{12}=\frac{16\pi\alpha_{s12}}{9m_1m_2}\delta({\bf r}_{12}),~
a_{34}=\frac{16\pi\alpha_{s34}}{9m_3m_4}\delta({\bf r}_{34}),~
a_{13}=\frac{32\pi\alpha_{s13}}{9m_1m_3}\delta({\bf r}_{13}),\\
a_{14}=\frac{32\pi\alpha_{s14}}{9m_1m_4}\delta({\bf r}_{14}),~
a_{23}=\frac{32\pi\alpha_{s23}}{9m_2m_3}\delta({\bf r}_{23}),~
a_{24}=\frac{32\pi\alpha_{s24}}{9m_2m_4}\delta({\bf r}_{24}).
\end{eqnarray}

Matrix elements with $\delta$-functions are expressed by the following formulas in terms 
of variational parameters:
\begin{equation}
\label{hfs3}
<\delta({\bf r}_{13})>=<\delta({\boldsymbol\sigma}+\frac{m_4}{m_{34}}{\boldsymbol\lambda}-
\frac{m_2}{m_{12}}{\boldsymbol\rho})>=\frac{8\pi^3}{(F_{1}F_{2}-F_{3}^2)^{3/2}},
\end{equation}
and the matrix elements of other delta functions $<\delta({\bf r}_{14})>$, 
$<\delta({\bf r}_{23})>$, $<\delta({\bf r}_{24})>$
are obtained from here by the same replacement
\eqref{f25}-\eqref{f27} as in the matrix elements of confinement potential. 
Remaining two matrix elements have the form:
\begin{equation}
\label{hfs4}
<\delta({\bf r}_{12})>=<\delta({\boldsymbol\rho})>=\frac{8\pi^3}
{(B_{22}B_{33}-B_{23}^2)^{3/2}},~
<\delta({\bf r}_{34})>=<\delta({\boldsymbol\lambda})>=\frac{8\pi^3}
{(B_{11}B_{33}-B_{13}^2)^{3/2}}.
\end{equation}

As numerical calculations show, in the case of identical quarks and antiquarks, the coefficients 
$a_{13}$, $a_{14}$, $a_{23}$, $a_{24}$ are equal in magnitude and equal to $a$. 
Then, when calculating hyperfine structure of the tetraquark spectrum, the following formula 
can be used:
\begin{equation}
\label{hfs6}
\Delta E^{hfs}=\frac{1}{2}a_{12}[s_{12}(s_{12}+1)-\frac{3}{2}]+
\frac{1}{2}a_{34}[s_{34}(s_{34}+1)-\frac{3}{2}]+\frac{1}{2}a[s_{T}(s_{T}+1)-
s_{12}(s_{12}+1)-s_{34}(s_{34}+1)].
\end{equation}

If pairs of quarks with spin 1 are considered, then the spin of the tetraquark takes the values 0, 1, 2, 
and corresponding energy values are equal to
\begin{equation}
\label{hfs7}
\Delta E(s_T=0)=\frac{1}{4}a_{12}+\frac{1}{4}a_{34}-2a,~
\Delta E(s_T=1)=\frac{1}{4}a_{12}+\frac{1}{4}a_{34}-a,~
\Delta E(s_T=2)=\frac{1}{4}a_{12}+\frac{1}{4}a_{34}+a.
\end{equation}

When adding up the spins of quarks, we use the following scheme: first, we add up the spins 
of the first and second $c$-quarks, the third and fourth $\bar c$-antiquarks. In this case, 
the total spins of these pairs are $S_{cc}=1$, $S_{\bar c\bar c}=1$ since the quarks are identical, 
and the antisymmetry of the wave function of the system is achieved due to its color part. 
Then we add up the two spins ${\bf S}_{cc}+{\bf S}_{\bar c\bar c}$ and we obtain that 
the total angular momentum of the tetraquark in the ground state takes the values 0, 1, 2. 
In this case, the charge and spatial parities of the tetraquark are equal to:
\begin{eqnarray}
\label{hfs8}
C_T=(-1)^{S_T+L_T},~~~P_T=(-1)^{L_T}.
\end{eqnarray}

\section{Conclusion}

In connection with new experimental data on the production of exotic bound states of quarks 
and gluons, a significant number of works have appeared in which the cross section calculation 
of the production of heavy tetraquarks in proton-proton interactions, in electron-positron annihilation, 
in B-meson decays is performed (see \cite{lebed} and references in it). 
To study such processes, it is necessary to calculate the wave function 
of the tetraquark and its static characteristics, the calculation of which is the subject of this work.
A distinctive feature of our calculations of the tetraquark mass spectrum is the use of the Gaussian wave 
functions in variational method.
We note that the results of calculating the tetraquark wave function obtained in this work have already 
been used by us in calculating the decay width of the Higgs boson with the production of a fully 
heavy tetraquark $(cc\bar c\bar c)$ \cite{tt}.

For the numerical calculation, a program is written in the MATLAB system for solving a four-particle 
problem within the stochastic variational method. The original Fortran program from \cite{v1} 
and our program for calculating three-particle energy levels from previous works \cite{p1,p2,p3}
are taken as a basis. Using trial Gaussian functions in the variational method, it is possible 
to obtain fairly accurate values of the energies of bound states of tetraquarks.

Numerical results of calculating the masses of ground states of tetraquarks taking into 
account the hyperfine structure are presented in Table ~\ref{tb1}. It also shows some results 
obtained by other authors in different models.

\begin{table}[htbp]
\caption{Masses of the ground states of tetraquarks in GeV}
\bigskip
\label{tb1}
\begin{tabular}{|c|c|c|c|c|c|c|c|c|} \hline
State &$(cc\bar c\bar c)$&\cite{efg3}&\cite{t2}&\cite{t12}&\cite{m1}&\cite{m2}& \cite{t13} &\cite{tc2} \\ \hline
$0^{++}$ &6.10 &6.190&6.477&5.966&6.797&5.883& 6.435  &  6.503    \\  \hline
$1^{+-}$ &6.26 &6.271&6.528&6.051& 6.899&6.120&  6.515  & 6.517       \\  \hline
$2^{++}$ &6.57 &6.367&6.573&6.223& 6.956& 6.246&  6.543 &  6.544      \\  \hline
\end{tabular}
\begin{tabular}{|c|c|c|c|c|c|c|c|c|c|} \hline
State &$(bb\bar b\bar b)$&\cite{efg3}& \cite{t12}&\cite{m1}&\cite{m2} & \cite{t13} \\ \hline
$0^{++}$ &18.81 &19.314 &18.754 & 20.155 &18.748 & 19.201   \\  \hline
$1^{+-}$ &18.86&19.320  &18.808 &20.212&18.828 &   19.251    \\  \hline
$2^{++}$ &18.97 &19.330 &18.916 &20.243 &18.900 &  19.262    \\  \hline
\end{tabular}
\begin{tabular}{|c|c|c|c|c|} \hline
State &$(cc\bar b\bar b)$ &\cite{efg3} & \cite{m1}&  \cite{m2} \\ \hline
$0^{++}$ &12.77&12.846&13.496 &12.445 \\  \hline
$1^{+-}$ &12.78 & 12.859&13.560&12.536  \\  \hline
$2^{++}$ &12.80&12.883& 13.595& 12.614 \\  \hline
\end{tabular}

\end{table}

Note that numerical results of the calculation change significantly with quark mass and
strong interaction constant.
Our choice of these parameters is the same as for mesons, although it remains an open question 
how the strong interaction constant might change when going from mesons to tetraquarks.
The maximum difference between the results of different authors is from 5 to 10 percent. 
We have taken into account relativistic corrections, and further refinement of the obtained 
results can be achieved by taking into account other corrections from the Breit Hamiltonian.
Although numerical results are obtained with high accuracy, we present our values with an accuracy
of two decimal places, since unaccounted corrections in the Hamiltonian of the system can contribute 
on the order of 0.1 percent.

\begin{acknowledgments}
The work was supported by the Foundation for the Development of Theoretical Physics and Mathematics 
BASIS (grant 22-1-1-23-1, 25-1-4-15-1). The authors are grateful to V.I. Korobov for useful discussions
and to Wen-Chao Dong for useful correspondence.

\end{acknowledgments}

\end{document}